# Designing Inclusive Crypto-Asset Dispute Resolution

## A Hybrid AI and Smart Contract Online Dispute Resolution Framework for Vulnerable Users

Gioia Arnone[1]*, Marco Giacalone[1]
[1]Department of Private and Economic Law (PREC), Vrije Universiteit Brussel
Brussels, Belgium – *Corresponding author: gioia.arnone@vub.be

## ABSTRACT

The growing use of crypto-assets has generated disputes that sit uneasily within existing legal redress mechanisms. Their resolution is complicated by the technical features of blockchain transactions, the cross-border nature of many relationships, and fragmented legal rules. These difficulties are particularly acute for users without legal or technical expertise, who may struggle to identify their rights, organise relevant evidence, or pursue an effective remedy. Crypto-assets may therefore produce new forms of digital exclusion, despite their association with financial accessibility. This paper examines whether a hybrid Online Dispute Resolution framework combining artificial intelligence and blockchain-based smart contracts could mitigate these barriers. In the proposed model, AI operates primarily off-chain, supporting natural-language interaction, dispute classification, evidence organisation, and accessible legal information. Smart contracts perform limited on-chain functions, including dispute registration, timestamping, verification, escrow management, and, where appropriate, execution of agreed outcomes. The framework is human-centred, with emphasis on explainability, procedural guidance, and human oversight in complex cases.

The model is examined through disputes involving crypto-asset seizure, exchange failures, and cross-border scams. The paper argues that a carefully delimited hybrid architecture may improve accessibility, transparency, efficiency, and enforceability, provided that automated and human roles are clearly defined. It also addresses key legal and governance concerns, including algorithmic bias, liability, data protection, smart-contract enforceability, and compliance with the EU AI Act and the Markets in Crypto-Assets Regulation. The paper thus proposes a legally grounded framework for more inclusive and robust dispute resolution in crypto-asset environments.

## CCS CONCEPTS

## KEYWORDS

## 1 INTRODUCTION

Crypto-assets have created a new category of disputes that does not fit comfortably within conventional legal and procedural frameworks. Conflicts may arise from fraudulent investment schemes, compromised wallets, irreversible payment errors, failures of trading platforms, or vulnerabilities embedded in decentralised finance protocols [1]. Resolving these disputes is especially difficult because the relevant transactions are often pseudonymous, technologically complex, and distributed across several jurisdictions. Existing remedies remain inconsistent and difficult to access, placing vulnerable individuals and users without specialist knowledge at a particular disadvantage [2].

The resulting problem extends beyond the mere absence of effective dispute resolution procedures. It reflects a broader form of digital inequality. Participation in crypto-asset markets may offer new financial opportunities, but meaningful access depends heavily on technical literacy, legal knowledge, and the resources required to obtain professional assistance. Users who lack these capabilities may be unable to reconstruct transactions, identify responsible parties, understand the applicable legal rules, or select an appropriate forum. Courts and traditional complaint mechanisms may also struggle with blockchain evidence, jurisdictional uncertainty, pseudonymous conduct, and the speed at which crypto-asset infrastructures evolve. These conditions contribute to a persistent access-to-justice deficit in the digital financial environment [3].

Emerging Legal Tech applications provide a possible basis for addressing this deficit. Artificial intelligence can help users describe and classify disputes, organise evidence, obtain preliminary legal guidance, and interact with procedural systems through natural-language interfaces. Blockchain-based smart contracts can support verifiable record-keeping, conditional asset management, and the execution of outcomes once predefined requirements have been fulfilled. Although both technologies have been examined extensively in isolation, comparatively little attention has been paid to their coordinated use in designing inclusive dispute resolution mechanisms for crypto-asset users [4].

This paper addresses that gap by developing a hybrid Online Dispute Resolution (ODR) framework in which AI-based legal support is combined with carefully limited smart-contract functions. The study asks: How can artificial intelligence and smart contracts be integrated into a dispute resolution architecture that improves digital inclusion and access to justice for vulnerable and non-expert users involved in crypto-asset disputes?

The research adopts a conceptual and design-oriented methodology. It proposes a legal and technical architecture that separates flexible, data-intensive, and human-centred activities from functions requiring verifiability and automated execution. AI tools operate mainly off-chain, assisting with user interaction, legal triage, dispute classification, evidence organisation, and plain-language explanations. The blockchain layer is reserved for selected functions, including timestamped dispute registration, integrity verification, escrow management, and the execution of validated settlement outcomes.

The proposed framework is built around accessibility, explainability, proportional automation, and human oversight. Guided workflows and simplified interfaces are intended to reduce cognitive and procedural burdens, while human mediators or legal professionals retain authority in cases requiring interpretation, discretion, or contextual assessment. Smart contracts are therefore not treated as substitutes for legal judgment, but as instruments supporting transparency, traceability, and reliable enforcement.

The model is examined through representative dispute scenarios involving asset seizure and recovery, exchange insolvency and mass claims, and scam-related cross-border losses. These examples reveal both the weaknesses of current remedies and the practical potential of a hybrid system linking accessible AI assistance to verifiable blockchain processes. The paper also considers the legal risks arising from such integration, including algorithmic bias, accountability, data protection, liability, and compliance with the EU regulatory framework governing AI and crypto-assets.

By bringing together legal accessibility, technical architecture, and automated enforcement, the paper contributes to the literature on AI and law with a structured model for inclusive digital justice in crypto-asset environments [5].

The paper proceeds as follows. Section 2 identifies the principal forms of crypto-asset disputes and the technical, legal, and cognitive dimensions of digital exclusion. Section 3 examines the role of AI in improving legal accessibility. Section 4 evaluates the use of smart contracts for verification and enforcement. Section 5 presents the hybrid ODR architecture, while Section 6 applies it to selected dispute scenarios. Section 7 addresses the relevant legal and regulatory implications. Section 8 proposes criteria for evaluating accessibility, fairness, efficiency, and enforceability. Section 9 concludes and outlines directions for future research.

## 2 BARRIERS TO JUSTICE IN CRYPTO-ASSET DISPUTES

The growing adoption of crypto-assets has produced a broad and increasingly complex range of legal conflicts. These disputes differ significantly in form and origin because they emerge from the distinctive technological, financial, and regulatory features of decentralised ecosystems. They may concern fraudulent schemes and scam-related losses, including Ponzi arrangements, phishing attacks, and deceptive investment offers; unauthorised transfers following the compromise of private keys; operational mistakes, such as irreversible payments sent to the wrong address; the collapse or insolvency of crypto-asset service providers; and failures within decentralised finance protocols, including defective smart contracts and governance-related conflicts. In contrast with conventional financial disputes, such cases often involve pseudonymous participants, irreversible transaction records, and digital infrastructures that are not clearly connected to a single jurisdiction [6].

These features make crypto-asset disputes particularly difficult to resolve through traditional legal and enforcement mechanisms. One of the most immediate problems is jurisdictional uncertainty. When a transaction involves actors, platforms, wallets, and technological infrastructure located across several countries, identifying the applicable law and the competent authority may become highly problematic. The absence of a clearly identifiable intermediary can make this determination even more difficult.

The treatment of evidence creates an additional obstacle. Blockchain ledgers may offer transparent, traceable, and tamper-resistant records, but the interpretation of wallet addresses, transaction hashes, token movements, and smart-contract interactions usually requires specialised technical knowledge. Such expertise is not readily available to many users, legal practitioners, or courts. Furthermore, the lack of harmonised standards for the presentation, authentication, and assessment of blockchain evidence complicates both judicial proceedings and alternative dispute resolution.

In this environment, digital exclusion must be understood in broader terms than the mere inability to access technology. It includes the technical, legal, and cognitive obstacles that prevent individuals from understanding their situation, exercising their rights, and participating effectively in dispute resolution. Barriers can be divided into three closely connected dimensions [7].

Technical exclusion derives from the complexity of blockchain-based systems. Users must manage wallets, private keys, seed phrases, cryptographic addresses, transaction hashes, and network fees, frequently through interfaces that are difficult to understand and offer little tolerance for error. Because many blockchain transactions cannot be reversed, even a minor mistake may lead to permanent financial loss. This risk is especially serious for non-expert users who may not know how to secure their assets, verify transaction details, or respond effectively when funds are stolen or misdirected [8].

Legal exclusion results from the fragmented and rapidly changing regulatory treatment of crypto-assets. Although several jurisdictions have introduced dedicated rules, including the European Union's Markets in Crypto-Assets Regulation, the legal framework remains uneven and difficult for ordinary users to interpret. The legal classification of a token, the identification of the responsible service provider, the recognition of consumer rights, and the availability of remedies may differ substantially from one jurisdiction to another. For individuals without legal training or access to professional advice, this uncertainty may prevent the effective pursuit of a claim [9].

Cognitive exclusion concerns the ability to understand the legal and technological information connected with a dispute. Relevant information may exist, but it is often expressed through specialised legal terminology, technical documentation, blockchain analytics, or opaque platform procedures. This lack of epistemic accessibility limits users' capacity to evaluate risks, understand available remedies, and make informed procedural choices. Some users may therefore abandon potentially valid claims, while others may pursue unsuitable or ineffective forms of redress [10].

Taken together, these dimensions reveal significant weaknesses in the existing mechanisms for resolving crypto-asset disputes. Traditional courts may be unable to respond efficiently to the technical complexity, cross-border character, and speed of such cases. Private dispute resolution services offered by exchanges or online platforms may be more accessible, but they can lack independence, transparency, accountability, or adequate procedural safeguards [11]. Consequently, many affected users remain without realistic or proportionate means of obtaining redress. This situation reinforces existing inequalities and may weaken confidence in crypto-asset markets and digital financial services.

Addressing these shortcomings requires a broader reconsideration of access to justice in decentralised environments [12]. It is not sufficient simply to apply existing legal procedures to technologically novel disputes. Dispute resolution systems must instead be designed around the specific characteristics of crypto-assets and the practical needs of different categories of users. From this perspective, digital inclusion should be treated as a multidimensional objective encompassing not only formal access to a procedure, but also usability, intelligibility, affordability, and the capacity to participate effectively [13]. The combined use of artificial intelligence and blockchain technology may provide a basis for developing such mechanisms [14].

AI systems can reduce informational and procedural barriers [15] by translating complex legal and technical material into accessible language, assisting users in identifying the nature of their dispute, structuring claims, organising evidence, and selecting suitable procedural options [16]. Blockchain-based smart contracts may complement these functions by supporting verifiable registration, conditional asset management, transparent record-keeping, and the execution of agreed outcomes.

Nevertheless, technology alone cannot guarantee inclusion. The effectiveness of an AI- and blockchain-supported dispute resolution system depends on the way it is designed and governed. Automation must remain proportionate, explainable, and subject to meaningful human oversight. Inclusive design must therefore constitute a central feature of the system rather than an additional or secondary consideration.

The following sections examine these components in greater detail. Section 3 considers how AI can improve legal accessibility and user participation, while Section 4 analyses the potential and limitations of smart contracts as instruments for verification and enforcement in crypto-asset dispute resolution ecosystems.

## 3 ARTIFICIAL INTELLIGENCE AS AN ENABLER OF ACCESSIBLE CRYPTO-ASSET DISPUTE RESOLUTION.

The increasing use of Artificial intelligence (AI) within legal services has created new opportunities to make legal information, procedural guidance, and dispute resolution more accessible. This potential is particularly relevant to crypto-asset disputes [17], where users must often confront complex technological systems, uncertain legal rules, and transactions extending across several jurisdictions. AI can help reduce these obstacles by assisting individuals who lack specialist legal or technical knowledge. Its capacity to promote digital inclusion [18], however, does not depend solely on technical performance. It also requires transparent design, appropriate governance, understandable outputs, and safeguards capable of protecting users from unfair or unreliable automated processes.

AI can support users throughout different stages of a crypto-asset dispute. At the initial stage, conversational systems and legal chatbots [19] can use natural language processing to interpret a user's account of events and provide preliminary guidance in clear and accessible terms. Such tools may help determine whether the dispute concerns fraud, an unauthorised transaction, the failure of a platform, or another type of crypto-related loss. They can also explain possible remedies and assist the user in deciding whether a formal dispute resolution procedure should be initiated [20]. By simplifying the first point of contact, these systems can reduce both informational barriers and the hesitation that often prevents non-expert users from pursuing a claim. It can also assist in organising and assessing disputes. Classification and decision-support tools may structure a claim, identify potentially relevant legal rules, and help users collect and arrange supporting evidence. In crypto-asset cases, these systems may analyse transaction histories, identify unusual movements of funds, recognise patterns associated with fraudulent conduct, and connect the facts described by the user with potentially applicable legal categories. Their role should remain supportive rather than determinative, since legal conclusions and disputed factual assessments require human judgment.

Nevertheless, they can improve the speed, consistency, and organisation of case handling [21].

A further application concerns negotiation and mediation. AI-assisted systems can support communication between parties, identify overlapping interests, suggest possible settlement options, and simulate the likely consequences of different solutions [22]. Within a hybrid ODR framework, these functions can complement the work of mediators by providing structured information and data-based insights without removing the flexibility and contextual assessment that human intervention provides [23]. This form of assistance may be especially valuable for vulnerable users because it can reduce disparities in knowledge, negotiating experience, and bargaining power [24].

Fraud detection and risk assessment tools may perform an additional diagnostic function. By examining transaction data and identifying recurrent behavioural patterns, AI can detect indicators associated with scams, money laundering, phishing, or other unlawful activities. In a dispute resolution setting, these systems may contribute to the preliminary validation of claims, the reconstruction of transaction pathways, and the identification of potentially responsible actors [25]. For users who are unable to interpret blockchain records independently, automated analysis can make complex technical information more understandable and can help them evaluate the credibility and seriousness of their claim. Taken together, these functions show that AI may provide support throughout the full dispute lifecycle, from the user's first interaction with the system to negotiation, assessment, and resolution [26]. Their contribution to inclusion, however, depends on whether they are embedded in accessible and genuinely user-centred processes.

Explainability is therefore a fundamental requirement. In disputes involving both legal and technological complexity, users must be able to understand how an AI system has reached a recommendation, classification, or assessment. This requirement is closely connected to epistemic accessibility, namely the ability of individuals to comprehend the information and procedures that influence their legal position and rights [27]. A system that produces conclusions without offering an intelligible explanation may be technically efficient but practically inaccessible.

Opaque or black-box systems may reinforce existing inequalities because users with advanced legal or technical knowledge are better able to question their outputs, while non-expert users may feel compelled to accept them without understanding their basis. For vulnerable individuals, this lack of transparency can reduce trust and limit meaningful participation in the dispute resolution process [28]. AI systems designed for legal inclusion should therefore explain relevant legal concepts in plain language, guide users through each procedural stage, clarify the reasons supporting their outputs, and openly communicate uncertainty, limitations, and the need for professional review. Explainability should not be regarded merely as an optional technical enhancement. It is closely linked to procedural fairness, accountability, and the user's ability to challenge or contest an automated assessment [29]. By making system reasoning more understandable, it strengthens user autonomy and promotes more informed engagement with digital justice mechanisms [30].

The use of AI also creates risks of discrimination and exclusion [31]. Bias may result from unrepresentative training data, unsuitable assumptions embedded in system design, inaccurate classifications, or the deployment of a model in a context for which it was not developed [32]. These concerns are particularly relevant in crypto-asset disputes, where reliable datasets may be scarce, fragmented, or disproportionately focused on specific transaction types, jurisdictions, or categories of users. An apparently neutral system may therefore produce less accurate or less favourable outcomes for groups that are poorly represented in the data.

Fairness must consequently be incorporated throughout the design and operation of the system. Training data should be reviewed for diversity and representativeness, model performance should be audited regularly, and outcomes should be examined for unjustified differences across user groups. Bias mitigation techniques may be necessary, but technical corrections alone are insufficient. Users should also have access to procedures through which they can provide feedback, challenge an output, request reconsideration, or seek human review [33]. Emerging regulatory frameworks increasingly require transparency, documented risk management, continuous monitoring, and meaningful human oversight in AI systems capable of affecting legal interests. Compliance with these standards is essential to ensure that digital dispute resolution does not reproduce the very inequalities it is intended to address [34].

Human involvement must therefore remain central to the proposed model. Human-in-the-loop mechanisms preserve the role of legal judgment, ethical reasoning, and contextual evaluation in cases that cannot be resolved through standardised rules [35]. Fully automated systems may process information quickly, but they may fail to recognise exceptional circumstances, power imbalances, vulnerability, or the broader consequences of a proposed outcome. These limitations are particularly significant where the facts are contested, the evidence is incomplete, or the dispute requires an equitable rather than purely rule-based solution.

A hybrid structure offers a more balanced allocation of responsibilities. AI may assist with routine functions, preliminary analysis, document organisation, and procedural guidance, while mediators, legal advisers, or other qualified professionals retain authority over decisions that affect substantive rights. Users should also be able to move from automated assistance to human support when the complexity or sensitivity of the case requires it. This arrangement combines efficiency and scalability with the safeguards associated with professional judgment and is consistent with regulatory approaches that require effective human control over high-impact AI applications.

The value of AI ultimately depends on whether users can interact with it in practice. A technically advanced system that is difficult to understand or navigate cannot meaningfully improve access to justice. User-centred design is therefore a necessary condition for digital inclusion [36]. Interfaces should avoid unnecessary legal and technical terminology, reduce cognitive burden, and guide users through clearly structured procedures. The system should accommodate different languages, devices, literacy levels, and accessibility needs. It should also explain what the AI can do, what it cannot do, and when human assistance is required.

These design choices are especially important in crypto-asset disputes because the underlying infrastructure is already highly complex. AI can perform an intermediary role between blockchain systems and users by translating transaction data, legal concepts, and procedural requirements into practical and understandable information. When developed according to principles of accessibility, transparency, fairness, and human oversight, AI can reduce cognitive, informational, and procedural barriers in crypto-asset dispute resolution [37]. Its effectiveness therefore depends not on automation alone, but on the creation of systems that remain intelligible, contestable, and responsive to human needs. The next section examines how smart contracts may complement these functions by supporting verifiable procedures and the reliable execution of dispute outcomes.

## 4 SMART CONTRACTS AS A TRUST AND ENFORCEMENT LAYER IN CRYPTO-ASSET DISPUTES

Within blockchain-based environments, smart contracts have become increasingly important as instruments for structuring and executing digital transactions. They consist of programmable code deployed on a distributed ledger and are designed to perform predetermined actions automatically once specified conditions have been satisfied [38].

In crypto-asset dispute resolution, this capacity may strengthen transparency, reduce transaction costs, and improve the implementation of agreed outcomes by limiting dependence on conventional intermediaries. At the same time, the use of smart contracts in legal processes introduces substantial technical, regulatory, and governance concerns, especially where the objective is to ensure access for vulnerable and non-expert users.

In a hybrid Online Dispute Resolution system, smart contracts may support several stages of the dispute process. One possible function is the secure registration and timestamping of a claim [39]. Rather than storing the complete dispute file on-chain, the system may record a cryptographic hash together with limited metadata, thereby generating tamper-resistant evidence that the claim existed at a particular moment. Such a mechanism can be particularly useful in cross-border disputes, where procedural rules and evidentiary requirements may differ between jurisdictions.

Smart contracts may also be used to manage disputed crypto-assets through escrow arrangements. Assets can be temporarily transferred to or controlled by a smart contract until the parties reach an agreement or a competent decision-maker determines the outcome. This reduces the risk that one party will withdraw, transfer, or conceal the assets while the dispute remains unresolved. By restricting unilateral control over the funds, escrow mechanisms may increase confidence in the process and discourage opportunistic conduct [40].

A further function concerns the implementation of settlements. Once the parties have reached an agreement through negotiation, mediation, arbitration, or another recognised procedure, the relevant obligations may be translated into programmable conditions. When those conditions are fulfilled, the smart contract can release funds, transfer assets, or perform another predefined operation without requiring separate enforcement proceedings. This capacity may narrow the gap that often exists between obtaining a decision and securing its practical implementation.

Blockchain infrastructure can also enhance procedural auditability. A carefully designed system may preserve an integrity-protected record of essential actions, including the registration of the dispute, the validation of an agreement, and the execution of the outcome. Such records can facilitate verification, strengthen accountability, and increase user confidence in digital dispute resolution mechanisms [41]. The value of this function lies not merely in immutability, but in the possibility of demonstrating that relevant procedural steps occurred in an identifiable and verifiable sequence.

These features are especially relevant in crypto-asset disputes because the parties may be pseudonymous, situated in different jurisdictions, or difficult to identify through conventional means. Traditional enforcement procedures can be costly, slow, and ineffective when the relevant assets are held in digital wallets and can be transferred rapidly. Automated execution may therefore improve the practical effectiveness of dispute outcomes, provided that the relevant assets remain within the technical reach of the smart contract.

From the perspective of digital inclusion, the automation of clearly defined enforcement functions may reduce the procedural and financial burden placed on users [42]. Individuals with limited resources may benefit from a process that does not require multiple applications, additional enforcement proceedings, or extensive professional assistance. Standardised smart-contract functions may also improve predictability by allowing users to understand in advance how an agreement will be implemented and what conditions must be fulfilled before assets are released.

Smart contracts may further reinforce trust in environments where conventional intermediaries are absent. Parties may be reluctant to participate in digital dispute resolution when they cannot be certain that the final agreement will be respected. A properly designed automated mechanism can provide assurance that an agreed outcome will be executed once the prescribed conditions are met. This does not eliminate the need for legal safeguards, but it can reduce dependence on the continuing cooperation of the opposing party.

When integrated with AI-based user interfaces, smart contracts may remain largely invisible to the user. The underlying code can operate as a technical enforcement layer, while the interface presents the process through understandable instructions, procedural explanations, and clearly defined choices. This separation is important because exposing users directly to code, wallet operations, and technical transaction logic would risk reproducing the same forms of exclusion that the ODR system is intended to address.

The use of smart contracts nevertheless presents serious limitations. Their most evident weakness is their rigidity. Smart contracts execute the instructions embedded in their code and generally cannot interpret ambiguous circumstances, reconsider unfair consequences, or respond appropriately to unexpected events. A system may therefore produce an outcome that is technically consistent with its programming but legally or substantively inappropriate. This risk is particularly significant in disputes requiring proportionality, equitable assessment, or consideration of individual vulnerability.

Legal recognition represents another unresolved issue. Although a smart contract can perform an operation on a blockchain, the legal significance of that operation may vary across jurisdictions. It may be unclear whether the code itself constitutes a binding contract, whether it merely implements an underlying legal agreement, or how coding errors and unintended outcomes should be addressed. Further uncertainty arises when the dispute concerns the smart contract itself, including questions of interpretation, validity, mistake, or defective performance. The absence of harmonised legal rules may therefore restrict the broader adoption of automated enforcement mechanisms.

Technical security constitutes an additional concern. Smart contracts may contain programming defects, design flaws, exploitable vulnerabilities, or unsafe dependencies on external systems. Their interaction with other contracts, digital wallets, data feeds, or blockchain protocols can produce unforeseen consequences. In decentralised finance, technical weaknesses have repeatedly resulted in significant financial losses. For non-expert users, who are rarely able to evaluate the security of code independently, such risks may intensify vulnerability rather than promote inclusion.

Questions of accountability and liability are equally important. Where a smart contract executes an incorrect or harmful transaction, responsibility may potentially lie with the developer, the platform operator, the person who supplied the relevant data, the party that approved the agreement, or another actor involved in the system. Determining which participant should bear responsibility may be difficult, particularly when development and operation are distributed among several entities. Effective remedies therefore require clear governance arrangements, traceable decision-making, and a defined allocation of responsibilities.

For these reasons, smart contracts should not be treated as autonomous substitutes for legal institutions or human judgment. Their more appropriate role is to support limited and clearly specified functions within a broader hybrid governance model. They may be used to register and verify procedural events, manage escrow arrangements, preserve integrity records, and execute settlement terms that have already been validated through an appropriate process. Interpretative, discretionary, and contested questions should remain subject to assessment by mediators, arbitrators, legal professionals, or other authorised human decision-makers.

This division of functions makes it possible to benefit from the speed, consistency, and reliability of automated execution while retaining the flexibility necessary for fair dispute resolution. Human oversight should be available before irreversible actions occur, particularly where the terms are disputed, the facts are uncertain, or execution may produce disproportionate consequences. Override, suspension, and review mechanisms are therefore essential components of a legally robust system.

A hybrid architecture can also support regulatory compliance by separating sensitive information and substantive decision-making from the blockchain layer. Personal data, evidence files, communications, and mediation records should generally remain off-chain, while only hashes, timestamps, validation records, and limited audit information are recorded on-chain. This structure reduces conflicts with data-protection requirements while preserving the evidentiary and integrity benefits associated with distributed ledgers.

The contribution of smart contracts to digital inclusion ultimately depends on their integration into accessible and user-centred ODR platforms. Users must receive clear information about the consequences of automated execution, the conditions governing the transfer or release of assets, and the circumstances in which human intervention is available. Technical complexity should be translated into understandable procedural steps, and the system should provide safeguards against coding errors, inaccurate data, and unintended outcomes.

Smart contracts can therefore improve the efficiency, transparency, and enforceability of crypto-asset dispute resolution, but only when their functions are carefully limited and embedded within a broader legal and institutional framework. Their value lies not in replacing human judgment, but in supporting verifiable records and reliable execution after the relevant legal and procedural questions have been resolved. The following section develops this approach by presenting a hybrid ODR architecture in which AI-based assistance, human oversight, and smart-contract enforcement operate as interconnected components of an inclusive digital justice system.

## 5 AN INCLUSIVE HYBRID ODR MODEL AND ITS APPLICATION TO CRYPTO-ASSET DISPUTES

The proposed framework translates the preceding analysis of digital exclusion, artificial intelligence, and smart-contract enforcement into a unified Online Dispute Resolution architecture designed for crypto-asset disputes. Its central purpose is to make technically and legally complex procedures more accessible to vulnerable and non-expert users while preserving transparency, procedural fairness, and the reliable execution of outcomes. The model combines off-chain AI-supported assistance and human decision-making with carefully limited on-chain functions, thereby creating a continuous process that connects user access, dispute assessment, negotiation, settlement, and enforcement.

The architecture is based on the principle that inclusion must be embedded in the design of the system rather than treated as an additional feature. User interaction should therefore take place through an accessible interface capable of presenting legal and technical information in clear language. Conversational tools may allow individuals to describe their dispute naturally, while guided procedures can reduce the difficulty of identifying relevant facts, uploading evidence, and selecting an appropriate remedy. Multilingual functionality, compatibility with different devices, and adaptation to varying levels of digital literacy are necessary to ensure that the platform does not exclude the users it is intended to assist.

Behind this interface, AI operates primarily off-chain as a cognitive and procedural support mechanism. It may classify the dispute, identify whether the case concerns fraud, an unauthorised transaction, platform insolvency, asset seizure, or another category of crypto-related harm, and assist in structuring the claim. AI tools may also organise documentary and blockchain-based evidence, explain transaction histories in plain language, detect indicators of suspicious activity, and provide preliminary guidance on possible procedural options. These outputs should remain explainable and open to review, particularly where they may influence the user's understanding of legal rights or the direction of the dispute.

The dispute resolution process itself occupies a hybrid decision space in which automated assistance is combined with human involvement. AI may help the parties identify areas of agreement, organise claims, compare proposed solutions, and model possible settlement outcomes. Nevertheless, mediators, legal advisers, arbitrators, or other competent professionals should retain responsibility for complex, contested, or legally significant determinations. Human oversight is particularly important where evidence is uncertain, the parties possess unequal bargaining power, or an automated recommendation may produce disproportionate consequences. The system is therefore designed to use automation as a form of assistance rather than as a substitute for contextual legal judgment.

Smart contracts provide the on-chain layer of the architecture and are reserved for functions that benefit specifically from verifiability, integrity, and conditional execution. A dispute may be registered through a timestamped cryptographic hash, creating proof that the claim and associated records existed at a particular moment without placing sensitive evidence directly on the blockchain. Where technically and legally appropriate, disputed assets may be held through an escrow mechanism until the parties reach an agreement or a recognised decision is issued. Once settlement terms have been validated, they may be translated into executable conditions so that the transfer, release, or allocation of assets occurs automatically when the agreed requirements are fulfilled. The blockchain may also preserve a tamper-resistant record of essential procedural events, thereby supporting auditability and trust.

The separation between off-chain and on-chain functions is fundamental to the model. User communications, personal information, evidence files, AI processing, negotiation, mediation, and substantive decision-making should remain off-chain, where they can be corrected, updated, restricted, or deleted in accordance with legal requirements. The blockchain layer should contain only limited information necessary to demonstrate integrity and execution, such as hashes, timestamps, validation records, access logs, and proofs of settlement. Secure interfaces connect the two environments by converting a legally or procedurally validated off-chain outcome into an on-chain condition for execution. This arrangement preserves the technical advantages of blockchain while limiting unnecessary exposure of personal or sensitive information.

In operational terms, the process begins when a user submits a complaint through the digital interface. The AI system then conducts preliminary triage, explains the nature of the claim, and assists in collecting and organising relevant evidence. The case is subsequently directed toward negotiation, mediation, or another appropriate procedure, with human intervention available whenever the dispute cannot be resolved through standardised assistance. When the parties reach a settlement, or when an authorised decision-maker establishes the outcome, the terms are reviewed and validated before any smart-contract execution occurs. The resulting agreement may then be implemented automatically, while a limited integrity record is retained for subsequent verification. This sequence establishes continuity between access, assessment, resolution, and enforcement without requiring the user to interact directly with the underlying technological infrastructure.

The model's practical value can be illustrated through disputes involving the seizure and recovery of crypto-assets. These cases often require the reconstruction of transaction pathways, the identification of wallets, the interpretation of blockchain evidence, and an understanding of the legal basis on which assets have been frozen or confiscated. Within the hybrid system, AI may translate transaction records into a comprehensible account of asset movements and provide users with accessible information about the relevant procedure and available remedies. The ODR environment may help structure the claim and coordinate the exchange of evidence among the user, legal professionals, service providers, and competent authorities. Where disputed assets remain technically controllable, an escrow mechanism may preserve them until a judicial or negotiated outcome determines their release. Such a process can reduce dependence on highly specialised expertise and enable affected users to participate more effectively.

The architecture is also relevant to disputes arising from the collapse or insolvency of crypto-asset service providers. These situations commonly involve large numbers of users, fragmented documentation, and complicated recovery procedures. AI-assisted guidance may standardise the submission of claims, identify missing information, and explain the user's procedural position. The ODR platform may aggregate similar claims, organise evidence, and support coordinated or collective resolution, thereby reducing administrative duplication. Once recovered assets become available for distribution, smart contracts may apply previously validated allocation rules and execute transfers in a transparent and consistent manner. Although such automation would remain subject to insolvency law and institutional approval, it could improve scalability and reduce procedural burdens in mass-claim settings.

Scam-related disputes and cross-border fraud provide a further field of application. Victims of phishing, fraudulent investment schemes, impersonation scams, or laundering networks frequently struggle to recognise the relevant transaction pattern, identify the responsible actors, and determine where to report or pursue the claim. AI tools may detect indicators associated with fraudulent behaviour, assist in reconstructing the sequence of transactions, and guide users through the preparation of a complaint. The ODR environment may facilitate cooperation among victims, platforms, investigators, mediators, and competent authorities across jurisdictions. Where assets have been traced and preserved, smart-contract mechanisms may support conditional restitution or controlled transfers following legal validation. This approach cannot eliminate the jurisdictional and enforcement difficulties associated with transnational fraud, but it may provide a more structured and accessible mechanism for coordinating claims and recovery efforts.

These applications show that the architecture can accommodate both individual and large-scale disputes. Across asset recovery, provider insolvency, and scam-related claims, AI reduces informational asymmetries by translating complex data and procedures into accessible forms, while the ODR environment provides a structured pathway for participation. Smart contracts contribute to enforcement where assets and outcomes can be translated into legally validated and technically executable conditions. Human oversight remains essential throughout the process to protect users from bias, factual error, rigid automation, and legally inappropriate outcomes.

The model nevertheless remains dependent on the quality of its implementation. Inaccurate data, poorly designed interfaces, unreliable AI outputs, insecure smart contracts, or unclear governance arrangements could reproduce or intensify exclusion. Users without basic connectivity or digital access may also remain outside the system. In addition, automated enforcement can operate effectively only when the relevant assets are technically reachable and the outcome is legally recognised. Continuous testing, regulatory adaptation, security assessment, and participatory evaluation are therefore required.

By combining AI-supported accessibility, human-centred dispute resolution, and limited blockchain-based enforcement, the proposed architecture links technological efficiency with the broader requirements of access to justice. Its significance lies not simply in placing different technologies within the same platform, but in assigning each component a proportionate and legally defensible role. AI supports understanding and participation, human actors preserve judgment and fairness, and smart contracts provide verifiability and execution where automation is appropriate. In this form, the hybrid ODR model offers a scalable and adaptable basis for addressing the technical, legal, and cognitive barriers that currently limit effective redress in crypto-asset environments.

## 6 REGULATORY ASSESSMENT OF THE HYBRID ODR MODEL COMPLIANCE AND PERFORMANCE

The practical implementation of a hybrid Online Dispute Resolution system for crypto-asset disputes requires more than a technically coherent architecture. Its legitimacy and effectiveness depend on whether it complies with the legal frameworks governing artificial intelligence, crypto-assets, personal data, contractual enforcement, and procedural justice, while also producing measurable improvements in accessibility, fairness, efficiency, and trust. Regulatory compliance and empirical evaluation should therefore be treated as interconnected elements of system design rather than as separate assessments conducted only after deployment.

Within the European Union, AI systems used in legal or dispute resolution contexts may be subject to heightened regulatory scrutiny, particularly when their outputs influence access to remedies or decisions with legal effects. AI tools used for legal triage, dispute classification, evidence assessment, or decision support may, depending on their function and context, fall within categories requiring enhanced risk management and oversight. The proposed model must consequently ensure that users are informed when they interact with an AI system, that automated outputs are sufficiently documented and understandable, and that system performance is monitored throughout its operational lifecycle. Meaningful human supervision must also remain available so that automated classifications, recommendations, or procedural decisions can be reviewed and corrected where necessary.

These obligations are consistent with the architecture's emphasis on explainability and human-in-the-loop governance, but they also create practical responsibilities for developers and platform operators. Compliance must be incorporated into the design process through documentation, testing, audit trails, risk assessments, performance monitoring, and clearly defined escalation procedures. An AI component should not be assessed only according to whether it produces technically accurate outputs. It must also be evaluated in relation to the legal consequences of those outputs, the ability of users to understand them, and the availability of effective mechanisms for contestation.

The system must also operate consistently with the regulatory framework applicable to crypto-asset markets. The Markets in Crypto-Assets Regulation establishes obligations for crypto-asset service providers relating to transparency, consumer protection, complaint handling, governance, and operational resilience [43]. Although the proposed ODR platform would not necessarily perform the functions of a financial intermediary, it could interact with exchanges, wallet providers, custodians, and other regulated entities. Its procedures should therefore reflect the rights and obligations that govern those actors, particularly in disputes involving inadequate disclosure, service failure, misconduct, insolvency, or loss of customer assets.

Integration with regulated entities may create opportunities to improve complaint management and supervisory cooperation. The platform could provide standardised channels for submitting claims, verifying procedural records, and communicating with relevant authorities. At the same time, such integration raises questions concerning institutional competence, jurisdiction, and the allocation of responsibility among platform operators, crypto-asset service providers, mediators, and public bodies [44]. These relationships must be governed by clear rules identifying who may access information, who may validate an outcome, and which institution remains legally responsible for the underlying procedure.

The legal effect of smart-contract execution constitutes another significant issue. A transaction may be performed automatically on a blockchain without necessarily resolving whether the underlying legal requirements of consent, intention, legality, capacity, and fairness have been satisfied. The status of code-based arrangements continues to differ across jurisdictions, and uncertainty remains concerning the treatment of programming defects, inaccurate inputs, unintended consequences, and disputes arising from the operation of the code itself.

For this reason, the proposed architecture does not treat smart contracts as independent substitutes for legal agreements or adjudicative decisions. Their function is primarily to implement outcomes that have already been established and validated through an appropriate human-mediated or legally recognised process [45]. This distinction reduces the risk that code will determine substantive rights without adequate legal scrutiny. Nevertheless, the off-chain legal agreement and its on-chain execution must remain closely aligned. The system must ensure that the programmed conditions accurately reflect the agreed terms and that an automated transfer or release of assets will be recognised as legally effective beyond the blockchain environment.

Accountability must be distributed clearly among the various participants involved in the system. AI developers, smart-contract programmers, platform operators, users, mediators, legal professionals, and external data providers may each influence the outcome of a dispute. Where an AI tool produces misleading guidance, responsibility may arise from defective training data, inappropriate model design, inadequate testing, or the failure to disclose limitations. Where a smart contract executes an incorrect transaction, the cause may lie in faulty code, erroneous data, insecure integration, or improper user authorisation. Where human review fails to prevent an unjust outcome, responsibility may also arise from deficient governance or procedural supervision.

A reliable accountability structure therefore requires traceability at every stage of the dispute process. The system should record the origin of relevant data, the basis of AI-supported recommendations, the identity and role of human reviewers, the validation of settlement terms, and the conditions triggering on-chain execution. The layered structure of the proposed model supports this objective by keeping substantive decision-making and sensitive information off-chain while recording limited integrity proofs and procedural events on-chain. Such documentation may assist both internal auditing and external regulatory review.

Data protection is equally central because ODR proceedings may involve identity information, wallet details, transaction histories, communications, allegations of fraud, and other sensitive material. Under the General Data Protection Regulation, personal data must be processed on a lawful basis and in accordance with the principles of data minimisation, purpose limitation, accuracy, security, and storage limitation. Users must also be able to exercise rights of access, correction, and, where applicable, erasure, while safeguards are required where automated processing significantly affects their legal position.

The permanence of blockchain records may conflict with these requirements, particularly where information must be corrected or removed. The architecture should therefore avoid placing personal data, evidence files, mediation communications, or substantive decisions directly on-chain. Instead, such material should remain in secure off-chain repositories, while the distributed ledger records only hashes, timestamps, verification data, and restricted audit information. This approach preserves the capacity to demonstrate integrity without unnecessarily exposing immutable personal information and may reduce the tension between blockchain technology and the right to erasure [46].

Compliance with formal regulation is not sufficient by itself. The system must also respect the fundamental requirements of procedural fairness and access to justice [47]. Users must have a meaningful opportunity to describe their claim, submit evidence, understand the procedure, respond to opposing arguments, and challenge an adverse recommendation or outcome. Decisions affecting substantive rights should not be delegated entirely to automated tools, and users should have access to an impartial human reviewer or decision-maker. Review, escalation, and appeal mechanisms are particularly important where the AI system expresses uncertainty, where the facts are contested, or where automated execution could have irreversible consequences.

These safeguards are also essential in cross-border disputes. Crypto-asset transactions may involve parties, service providers, assets, and technological infrastructures located in several jurisdictions. Although a digital platform can provide a common procedural entry point, it cannot by itself resolve all questions of applicable law, jurisdiction, recognition, and enforcement. The legal effect of an ODR settlement or decision may differ between countries, and interaction with national courts or regulatory authorities may remain necessary. The architecture should therefore be capable of adapting its procedures to jurisdiction-specific requirements while maintaining common standards of transparency, accessibility, and record integrity.

The system's regulatory legitimacy must be supported by systematic evaluation of its real-world performance. A conceptually inclusive platform cannot be assumed to improve access to justice merely because it uses accessible interfaces, AI assistance, or automated enforcement. Its operation must be tested against measurable legal, technical, and user-centred criteria. Evaluation should determine whether vulnerable and non-expert users can access the platform, understand the information provided, participate effectively, obtain fair outcomes, and secure reliable enforcement.

- Accessibility should be examined in relation to the practical ease with which users can enter and navigate the platform. This includes compatibility with different devices, connectivity conditions, languages, disabilities, and levels of digital competence. A system may be available online while remaining inaccessible in practice because its interface is too complex, its requirements assume advanced technical knowledge, or its functions cannot be used through commonly available devices. Accessibility must therefore be evaluated independently rather than inferred from the mere existence of a digital service.
- Usability and comprehension should assess whether users can understand the AI-generated explanations, complete the guided procedures, identify the implications of different choices, and act on the information provided. Comprehension testing is especially important because an interface may appear clear to developers or legal professionals while remaining confusing to non-expert users. One relevant indicator is the proportion of participants who can accurately explain their legal position and the next procedural step after interacting with the system.
- Fairness requires an examination of whether classifications, recommendations, and outcomes differ unjustifiably across user groups. The performance of AI tools should be compared across relevant demographic, linguistic, socioeconomic, and digital-literacy categories, while recognising the legal and ethical limits applicable to the collection and use of such data. The evaluation should also determine whether users can challenge automated outputs, obtain human review, and receive a reasoned response. Fairness cannot be established solely through statistical consistency; it also depends on procedural opportunities for correction and contestation.
- Efficiency should be assessed by comparing the hybrid ODR process with existing judicial, administrative, or platform-based procedures. Relevant measures include the average time between complaint submission and resolution, the number of procedural steps required, the cost incurred by users, and the resources required from mediators or platform operators. Reduced processing time and lower costs may improve access to justice, but efficiency should not be pursued at the expense of accuracy, fairness, or meaningful human supervision.

- Enforceability and trust should be evaluated by examining whether agreed outcomes are implemented correctly and whether the underlying records remain complete, secure, and verifiable. The reliability of smart-contract execution may be measured through the proportion of validated settlements executed without error, delay, or unauthorised intervention. User trust should also be assessed directly, since technical reliability does not necessarily produce confidence where users do not understand how the system operates or who is responsible when problems occur.

These dimensions can be translated into measurable indicators, including user comprehension rates, average resolution times, comparative cost reductions, differences in outcomes across user groups, satisfaction levels, rates of successful human escalation, and the reliability of automated execution. The resulting data should be interpreted alongside qualitative evidence gathered from interviews, observation, and user feedback. A mixed-methods approach is particularly suitable because it combines measurable performance indicators with insight into how users experience and understand the procedure [48].

Evaluation should take place through user testing, realistic dispute simulations, case-based analysis, security audits, and continuous monitoring of AI and smart-contract performance. Participants should include non-expert crypto-asset users, individuals with limited digital literacy, legal practitioners, mediators, technical specialists, consumer representatives, and relevant institutional actors. Involving these groups can reveal barriers that would remain invisible in a purely technical assessment.

The process should be iterative rather than limited to a single validation exercise. AI models, blockchain protocols, user behaviour, and regulatory requirements will continue to evolve. Continuous monitoring and feedback mechanisms are therefore necessary to detect emerging bias, security weaknesses, usability problems, and compliance failures. Evaluation results should lead to concrete modifications in system design, governance arrangements, explanatory content, and escalation procedures.

Regulatory compliance and performance assessment are thus mutually reinforcing. Legal standards establish the safeguards that the system must provide, while empirical evaluation determines whether those safeguards operate effectively in practice. A platform may formally include explainability, human oversight, and review mechanisms yet fail to make them meaningful or accessible to users. Conversely, empirical evidence of strong performance cannot justify disregard for mandatory legal protections. The hybrid ODR model can promote digital inclusion only when legal legitimacy, technical reliability, and user experience are assessed together.

The ultimate question is not simply whether the architecture functions as intended, but whether it reduces the structural obstacles that prevent users from obtaining effective redress. A successful system should enable individuals to understand their legal position, participate meaningfully in the procedure, challenge automated outputs, receive fair treatment, and obtain an enforceable outcome. By combining regulatory safeguards with continuous and participatory evaluation, the proposed framework provides a basis for determining whether AI and smart contracts genuinely advance access to justice in crypto-asset disputes rather than merely digitalising existing inequalities.

## 7 CONCLUSION AND FUTURE DIRECTIONS

Crypto-asset disputes reveal a growing mismatch between the rapid development of decentralised financial environments and the ability of existing legal institutions to provide accessible, timely, and effective remedies. The technical opacity of blockchain transactions, the fragmentation of applicable legal rules, the pseudonymity of participants, and the cross-border movement of digital assets place particularly significant burdens on vulnerable and non-expert users. As this paper has demonstrated, these difficulties cannot be addressed merely by transferring conventional dispute resolution procedures to an online environment. Effective digital justice requires a system designed around the specific characteristics of crypto-assets and the practical capacities of those seeking redress.

The proposed hybrid ODR architecture responds to this challenge by assigning distinct but interconnected roles to artificial intelligence, human decision-makers, and smart contracts. AI operates primarily as an off-chain assistance mechanism, supporting legal triage, dispute classification, evidence organisation, fraud detection, procedural guidance, and the translation of complex information into accessible language. Human professionals retain responsibility for contested facts, legal interpretation, equitable assessment, and decisions affecting substantive rights. Smart contracts perform narrowly defined on-chain functions where verifiability and conditional execution provide a genuine advantage, including timestamped registration, integrity verification, escrow management, and the implementation of previously validated settlements.

This allocation of functions is central to the model's contribution. AI is not presented as an autonomous legal decision-maker, nor are smart contracts treated as substitutes for courts, mediators, or legally recognised agreements. Instead, the architecture uses automation proportionately, preserving human oversight before irreversible or legally consequential actions occur. Its off-chain and on-chain separation also limits the storage of personal and evidentiary information on immutable ledgers, thereby supporting data minimisation, correction, and erasure while retaining the integrity benefits of blockchain verification.

The application of the framework to asset recovery, service-provider insolvency, mass claims, and cross-border scams demonstrates its potential relevance across different categories of crypto-asset disputes. In each context, AI-supported interfaces can reduce informational and cognitive asymmetries, structured ODR procedures can improve participation and coordination, and smart contracts can strengthen execution where assets remain technically accessible and the outcome has been legally validated. Nevertheless, the architecture cannot eliminate every obstacle associated with crypto-asset disputes. Jurisdictional fragmentation, difficulties in identifying responsible actors, the legal recognition of digital outcomes, insecure code, unreliable data, and the exclusion of individuals without adequate digital access remain substantial limitations.

The effectiveness of the model therefore depends on regulatory compliance, institutional cooperation, and continuous empirical assessment. Explainability, accountability, data protection, human review, and procedural contestability must be embedded throughout the system's lifecycle. Its performance should be evaluated not only through technical measures, but also by examining whether users can understand their legal position, participate meaningfully, obtain fair treatment, and secure an enforceable remedy. Mixed-methods evaluation combining usability testing, realistic case simulations, performance auditing, stakeholder participation, and analysis of user outcomes provides an appropriate basis for determining whether the system advances inclusion in practice.

The paper's principal contribution lies in showing that accessibility and enforceability should not be treated as competing objectives. A carefully governed hybrid architecture can use AI to make dispute procedures more intelligible, human oversight to preserve fairness and contextual judgment, and smart contracts to provide integrity and reliable execution. Yet these benefits depend on design choices rather than on the technologies themselves. Without accessible interfaces, transparent reasoning, meaningful review, and clear responsibility structures, digital dispute resolution may simply reproduce existing inequalities in a more technologically sophisticated form.

Future research should move beyond conceptual design towards empirical validation through prototype development, controlled simulations, user testing, and comparative studies across jurisdictions. Particular attention should be given to the experiences of individuals with limited digital literacy, the accuracy and explainability of AI-generated guidance, the security of smart-contract execution, and the interaction between ODR outcomes and national enforcement systems. Such research will be essential to determine the conditions under which the proposed framework can be implemented responsibly and at scale.

Ultimately, inclusive crypto-asset dispute resolution requires more than technological innovation. It demands a socio-technical and legally grounded approach in which accessibility, procedural fairness, accountability, and enforceability are incorporated from the outset. By integrating AI-supported assistance, human judgment, and limited blockchain-based execution within a coherent governance structure, the proposed model provides a foundation for transforming crypto-asset dispute resolution from a fragmented and exclusionary process into a more understandable, trustworthy, and effective form of digital justice.